\documentclass{cwauth}

\usepackage{moreverb}

\usepackage{times}

\usepackage[colorlinks,bookmarksopen,bookmarksnumbered,linkcolor=blue,citecolor=blue,urlcolor=blue]{hyperref}

\usepackage{array}
\usepackage{multirow}
\usepackage{amsmath}
\usepackage{amssymb}
\usepackage{graphicx}

\providecommand{\tabularnewline}{\\}

\begin{document}

\runninghead{C.W. et al.}

\title{A multi-locus predictiveness curve and its summary assessment for genetic risk prediction}

\author{Changshuai Wei\affil{a}\corrauth,
Ming Li\affil{b}, Yalu Wen\affil{c}, Chengyin Ye\affil{d} and Qing Lu\affil{e}\corrauth}

\address{\affilnum{a} Department of
Biostatistics and Epidemiology, University of North Texas Health Science Center.\\
\affilnum{b} Department of Epidemiology
and Biostatistics, Indiana University Bloomington.\\
\affilnum{c} Department of Statistics, University of Auckland\\
\affilnum{d} School of Medicine, Hangzhou Normal University.\\
\affilnum{e} Department of Epidemiology and Biostatistics, Michigan State University.}

\corraddr{Qing Lu, Department of Epidemiology and Biostatistics, Michigan State University. E-mail: qlu@epi.msu.edu. \\
Changshuai Wei, Department of
Biostatistics and Epidemiology, University of North Texas Health Science Center. E-mail: changshuai.wei@unthsc.edu\\
This is the author's manuscript of a paper published in \href{https://doi.org/10.1177/0962280218819202}{https://doi.org/10.1177/0962280218819202}}

\begin{abstract}

With the advance of high-throughput genotyping and sequencing technologies, it becomes
feasible to comprehensive evaluate the role of massive genetic predictors in disease prediction.
There exists, therefore, a critical need for developing appropriate statistical measurements to
access the combined effects of these genetic variants in disease prediction. Predictiveness curve
is commonly used as a graphical tool to measure the predictive ability of a risk prediction model
on a single continuous biomarker. Yet, for most complex diseases, risk prediciton models are
formed on multiple genetic variants. We therefore propose a multi-marker predictiveness
curve and provide a non-parametric method to construct the curve for case-control studies. We further introduce a global predictiveness U and a partial predictiveness U to summarize
prediction curve across the whole population and sub-population of clinical interest,
respectively. We also demonstrate the connections of predictiveness curve with ROC curve and Lorenz curve. Through simulation, we compared the performance of the predictiveness U to
other three summary indices: R square, Total Gain, and Average Entropy, and showed that
Predictiveness U outperformed the other three indexes in terms of unbiasedness and
robustness. Moreover, we simulated a series of rare-variants disease model, found partial
predictiveness U performed better than global predictiveness U. Finally, we conducted a real
data analysis, using predictiveness curve and predictiveness U to evaluate a risk prediction
model for Nicotine Dependence.
\end{abstract}

\keywords{High-dimensional Data; Non-parametric Statistic; Nicotine Dependence}

\maketitle

\section*{Introduction }

Genome-wide association studies have discovered thousands of disease-susceptibility loci
for complex diseases. Although each locus has only a moderate or low effect on diseases, jointly
they can significantly influence the disease risk. Risk predictive models capitalizing on the
combined information of newly identified genetic variants and existing risk predictors thus hold
great promise for individualized disease prediction and prevention. For this reason, studies to
assess the combined role of newly genetic loci in early disease prediction represent high priority
research projects, as manifested in the multiple risk predictive modeling studies now underway. The yield from these ongoing risk prediction studies can be enhanced by adopting appropriate
methods and measurements that comprehensively evaluate risk prediction models, including assessing classification accuracy of models and their contribution to the absolute
risks on the population scale.

The receiver operating characteristic (ROC) curve have been a very popular
tool to evaluate clinical validity of a risk prediction model. It comprises of all possible pairs of
senesitivty and specificity, and there is a global measure of a model's classficiation accucy. Yet, the
ROC curve does not assess the clinical utility of the model.  To access the prediction performance of
a risk model and its clinical utility on population level, Pepe et al \cite{Pepe2008} propose a graphical
tool, predictiveness curve. Predictiveness
curve is constructed by plotting each
individual's predicted risks versus its quantile,
or in another word, the transpose of the cumulative distribution
function of risk distribution.
Thus, the predictiveness curve displays the information of the risk
distribution on population level. Huang et al \cite{Huang2007,Huang2010}studies the asymptotic property
of the predictiveness curve and developed methods for estimating the
predictiveness curve of single continuous biomarker for case-control
design. Nevertheless, complex diseases are commmonly manifested by mutliple risk predictors, which call the need of predictiveness curve for mutliple risk predictors.

Besides graphical evaluation, a summary index for predictiveness curve is necessary for numerically evaluation of prediction models, especially when the number of models for comparison is large. Several summary indexes have been proposed for predictiveness curve of a single continuous marker, for instance, the variance of the predicted risks. Intuitively, a large variance of predicted risk
means the marker assigned different risks to subjects, indicating
good prediction performance of the marker. If we divide
the variance by the variance of a perfect predictor, we obtain a R-square
statistic\cite{Mittlbock1996}. Considering
R square is not suitable for a binary response\cite{Cox1992}, two other summary indexes were proposed: Average Entropy and Total Gain.\cite{Huang2007,Shapiro1977,Bura2001}
Average Entropy was defined by calculating the reduction in maximum log-likelihood, and can be considered a generalized R-square statistic. Total Gain was proposed by Bura et al\cite{Bura2001} to summarize the binary regression quantile plot. Total Gain can be also considered as a generalized R-square
statistic, by replacing the squared risk difference with the absolute
value of risk difference. The existing summary indexes, however, were developed for single continuous predictor, where there is a natural underlying order. For multiple predictors, the order is yet to be defined and summary indexes without considering ordering information may not have optimum performance.

U statistic\cite{Hoeffding1948} is well known for its flexibility and robustness, and have been recently applied to genetic association study by incorperating ordering information of multiple genetic variants\cite{Wei2008,Schaid2005}. U statistic, in particular, Mann-Whiteney U, has been shown to be an optimum summary index for ROC curve\cite{Lu2008}. Thus we proposed a predictiveness U to
summarize the predictiveness curve for multiple predictors with consideration of ordering information. In the following sections, we first developed a predictiveness curve and predictiveness U for multiple predictors. We then showed their connections with ROC curve and Lorenze curve. We provided asymptotic results for predictiveness U for efficient inference. At last, we performed simulations to compare predictiveness U with other summary indexes followed with a real data application on Nicotine Dependence.

\section*{Method }

\subsection*{Multimarker predictiveness curve}

Assume we have an random variable $\mathcal{G}$, which represents
multiple genetic variants. Let $F(g)$ be the cumulative distribution
function (c.d.f ) of $\mathcal{G}$, such that $F(g)=P(\mathcal{G}\leq g)$.
Then, the predictiveness curve can then be defined as a function that maps genetic quantile $q=F(g)$ to predicted risk $r(q)$,

\[
r:q\rightarrow r(q).
\]
It is crucial to order the genetic
varaints $g$, so as to have meaningful definition of $F(g)=P(\mathcal{G}\leq g)$.
Here, we assume that the order of $g$ is concordant with the order
of predicted genetic risk $r$, so that the predictiveness curve is
an monotone increasing curve. In fact, under the monotone increasing
assumption, predictiveness curve is just the inverse cumulative distribution
function of predicted genetic risk $r=F_{r}^{-1}(q)$, where $F_{r}$
is c.d.f of $r$.

Because there is natural order underlying a single continuous marker, it is convenient to obtain a monotonically increasing predictiveness curve for the single marker. However, for multiple markers, especially the
categorical genetic variants, the underlying order is unknown. Nonetheless,
we perform transformation on the categorical multi-locus genotypes, making
the corresponding predictiveness curve monotonically
increasing with the transformed genotype. To
illustrate, we assume there are $m$ genetic markers in the risk prediction
model, which comprise of $G_{m}$ multi-locus genotypes. Let  $r_{i}$ ($r_{i}=P(D|g_{i})$) and $p_{i}$ ($p_{i}=P(g_{i})$ ) denote the predicted risk and the population proportion of the multi-locus genotype $g_{i}$ ($1\leq i\leq G_{m}$). Given a binary disease phenotype with the disease prevalence of $\rho$, we can calculate $r_i$ and $p_i$:
\[
r_{i}=\frac{P(g_{i}|D)\rho}{P(g_{i}|D)\rho+P(g_{i}|\bar{D})(1-\rho)},
\]
\[
p_{i}=P(g_{i}|D)\rho+P(g_{i}|\bar{D})(1-\rho),
\]
where $P(g_{i}|D)$ and $P(g_{i}|\bar{D})$ denote proportion of $g_i$ carrier in diseased sub-population and  non-diseased sub-population, respectively. We
ordered the multi-locus genotype from lowest risk to highest risk
(so that $r_{i}>r_{j}$ for$i>j$) and plotted the ordered risk ($r_{i}$
) against the risk quantile $q_{i}$ ($q_{i}=\sum_{j=0}^{i}p_{i}$),
whereby we obtain a multi-marker predictiveness curve. 

For case-control data, $P(g_{i}|D)$ can be estimated as $\hat{P}(g_{i}|D)=\frac{n_{D,g_{i}}}{n_{D}}$,
where $n_{D}$ is the number of cases, and $n_{D,g_{i}}$ is the number
of $g_{i}$ carrier in cases. $\hat{P}(g_{i}|\bar{D})$ can be estimated 
in a similar way. The disease prevalence $\rho$ can be obtained
from previous literature or another independent cohort study. With $\hat{P}(g_{i}|D)$, $\hat{P}(g_{i}|\bar{D})$
and $\rho$ , we can obtain $\hat{r}_{i}$, $\hat{p_{i}}$ and thus
an estimator of the predictiveness curve.

\subsection*{Predictiveness U}
Various single marker summary indexes have been proposed to summarize predictiveness curve. Most of them, however, are not approporate for predictiveness curve built on multiple
markers. To measures the prediction performance of a set of markers on the population level, we propose a predictiveness U.
Given the predictiveness curve, we can calculated the predictiveness U by,
\[
U=2\sum_{i>j}p_{i}p_{j}\psi(r_{i},r_{j}),
\]
where the kernel $\psi(r_{i},r_{j})$ measures the difference of disease
impact between $g_{i}$ carrier and $g_{j}$ carrier, $p_{i}p_{j}$
weights the impact on the population level. Different forms of $\psi(r_{i},r_{j})$ can be chosen so as to incorporate
cost associated with disease risks. Here in this paper, we use
risk difference $\psi(r_{i},r_{j})=r_{i}-r_{j}$.

The above definition can be generalized to the
general predictiveness curve, $r:q\rightarrow r(q)$, where we define the predictiveness U as 
\[
U=2\int_{0}^{1}\int_{0}^{y}\psi(r(y),r(x))dxdy.
\]
When $\psi(r_{i},r_{j})=r_{i}-r_{j}$ , the
predictiveness U can be written as $U=2\int_{0}^{1}\int_{0}^{y}(r(y)-r(x))dxdy$.
Furthermore, we can standardize predictiveness U by dividing it by
its maximum value (i.e., perfect prediction). For risk difference
kernel, the standardized predictiveness U can be calculated by $U^{st}=[\int_{0}^{1}\int_{0}^{y}(r(y)-r(x))dxdy]/[\rho(1-\rho)]$,
which scale from 0 to 1.

In a population of size $N$, suppose genotype $g_{i}$ has
$N_{i}$ carrier and is associated with disease risk $r_{i}$. With $p_{i}=N_{i}/N$,
the Predictiveness U can be written as a one sample U-statistic,
\[
U=\frac{\sum_{i>j}N_{i}N_{j}\psi(r_{i},r_{j})}{\frac{N(N-1)}{2}}
\]
Using Hoeffding projection, we can calculate
the population variance of predictiveness U by:

\[
var(U)=\frac{4}{N}var(E(\psi(r_{1},r_{2})|r_{1})=\frac{4}{N}\sum_{i=1}^{G_{m}}\frac{N_{i}}{N}(\sum_{j=1}^{G_{m}}\frac{N_{j}}{N}\psi(r_{i},r_{j})-U)^{2}.
\]

In clinical application, our interest may lie in a sub-population with specific range of risk quantile $q\in(q_{0},q_{1})$.
We can define a partial predictiveness U, $U_{pt}=2\int_{q_{0}}^{q_{1}}\int_{q_{0}}^{y}\psi(r(y),r(x))dxdy$, to evaluate the prediction performance of a model on the sub-population.
With the given risk difference kernel, the partial predictiveness U becomes,
\[
U_{pt}=2\int_{q_{0}}^{q_{1}}\int_{q_{0}}^{y}(r(y)-r(x))dxdy.
\]
We can also standardize partial predictiveness
U to make it ranges 0 to 1, i.e., $U_{pt}^{st}=[\int_{q_{0}}^{q_{1}}\int_{q_{0}}^{y}(r(y)-r(x))dxdy]/[\rho_{pt}(1-\rho_{pt})]$,
where $\rho_{pt}=\int_{q_{0}}^{q_{1}}r(x)dx$. The calculation of the population variance for partial predictiveness U is similar as that of global predictiveness U by considering constraint on the selected range of risk quantile.

\subsection*{Connection with ROC curve and Lorenze Curve}

\begin{figure}
\protect\protect\caption{Predictiveness curve, ROC curve and Lorenz curve}
\label{Fig1}

\centering \includegraphics[scale=0.52]{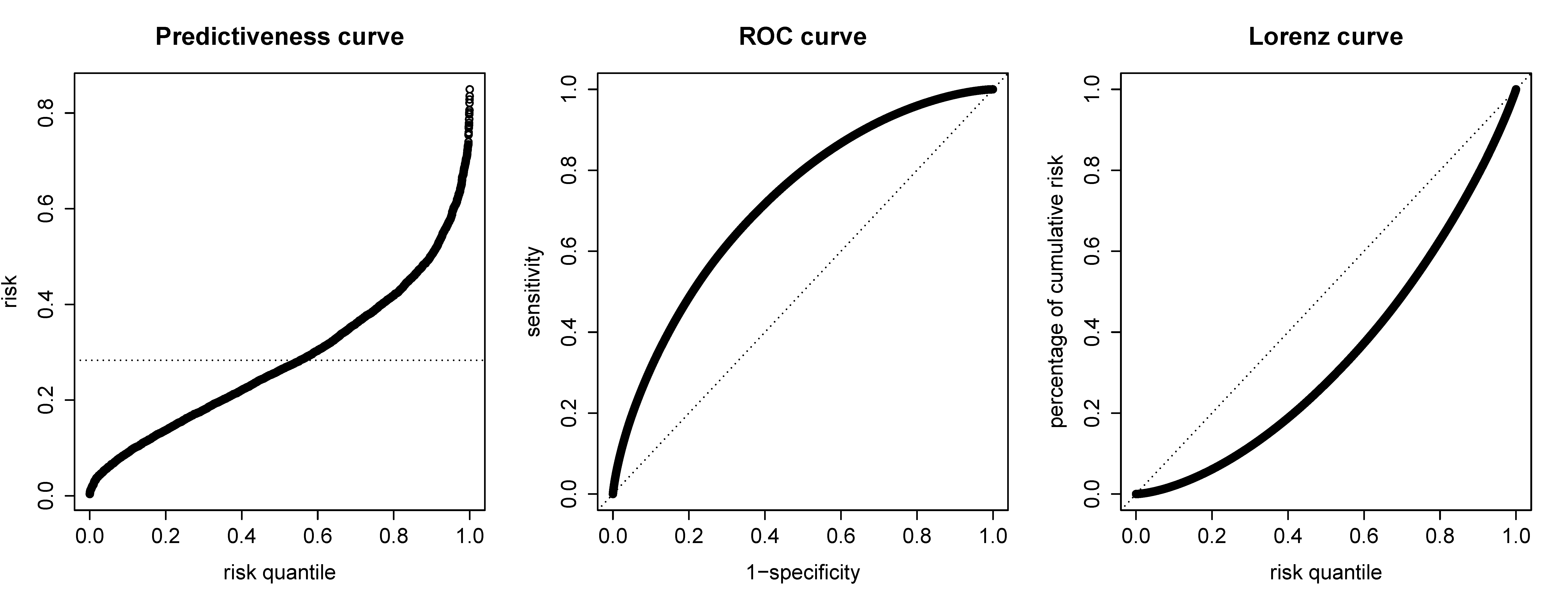}
\end{figure}

In this subsection, we show the connection of predictiveness
curve and ROC curve and Lorenze Curve (Figure \ref{Fig1}). For simplicity, we
limit our discussion to predictiveness U with risk difference kernel
$\psi(r_{1},r_{2})=r_{1}-r_{2}$.

The ROC curve has been a popular visualization method to evaluate the classification accuracy of a genetic
risk prediction model\cite{Lu2008}. It is formed by plotting the
$sensitivity$ against $1-specificity$ at various thresholds. The area under the ROC curve
$AUC_{R}$ is often used as a summary index for classification accuracy of the prediction model, where higher value of $AUC_{R}$ represents higher classification accuracy. With a threshold genotype $g$, we can have the following decision rule for the predicted
disease status $Y_g$,
\[
Y_g=\begin{cases}
1, & \mathcal{G}>g,\\
0, & \mathcal{G}\leq g.
\end{cases}
\]
The ROC curve can then be formed by plotting $sensiticity$ (i.e., $P(Y_g=1|D)$ against $1-specificity$ (i.e., $1-P(Y_g=0|\bar{D})$) at various threshold $g$. In particular, the ROC curve is represented by a map: $f:t\rightarrow f(t)$, where
$t_g=1-P(Y_g=0|\bar{D})$ and $f(t_g)=P(Y_g=1|D)$.
Let $F_{D}(g)$, and $F_{\bar{D}}(g)$ denote the c.d.f of ordered
genotype $\mathcal{G}$ in the diseased sub-population and non-diseased sub-population. With
decision rule on $\mathcal{G}$, the ROC curve can be represented as, 
\begin{align*}
t_g & =1-F_{\bar{D}}(g),\\
f(t_g) & =1-F_{D}(g).
\end{align*}
Meanwhile, the predictiveness curve can be written as (Appendix A),
\begin{align*}
q_g & =\rho F_{D}(g)+(1-\rho)F_{\bar{D}}(g),\\
r(q_g) & =\frac{F_{D}^{\prime}(g)\rho}{F_{D}^{\prime}(g)\rho+F_{\bar{D}}^{\prime}(g)(1-\rho)}.
\end{align*}

Thus both ROC curve and predictiveness curve can be written as function
of $F_{D}(g)$ and $F_{\bar{D}}(g)$. Further, we can show (Appendix A)
that $AUC_{R}$ and predictiveness U have the following
relationship,

\[
U=2\rho(1-\rho)(2AUC_{R}-1).
\]

Since the diagnal line for ROC curve represents the discriminative ability of
a non-informative marker, the area between ROC curve and diagnal line
($\Delta AUC_{R}=AUC_{R}-0.5$) represents the net classifcation improvement
of a predictive marker. In other words, the predictiveness U is just the net classification improvement
of the ROC curve multiplied by a factor related to disease prevalence
, i.e., $U=4\rho(1-\rho)\Delta AUC_{R}$.

Lorenz curve was originally invented to measure the income inequality. Nonetheless, We can also use Lorenz curve to measure risk inequality
in the population. Each point on the Lorenz curve maps the risk quantile
$q$ to the percentage of cumulative risk $h(q)=\frac{\int_{0}^{q}r(u)du}{\rho}$.
The diagonal line in Lorenz curve represents the scenario of no risk
inequality, where all the individuals in the population have the same
risk. In a real population, the Lorenz curve should be an convex curve.
The further away the Lorenz curve is from the diagonal line, the more
risk inequality the population exhibits. Denote the area under the Lorenze curve as $AUC_{L}=\int_{0}^{1}h(u)du$.
The degree of risk inequality can be measured by the area between
the diagnal line and the Lorenze curve, $\Delta AUC_{L}=0.5-AUC_{L}$.
We can show (Appendix B) that $AUC_{L}$ and predictiveness U have the following
relationship 
\[
U=2\rho(0.5-AUC_{L}),
\]
which means that predictiveness U can be represented as the degree of
risk inequality multiplied by a factor related to disease prevalence, i.e., $U=2\rho\Delta AUC_{L}$.

\subsection*{Estimation and inference}

We can apply an existing prediction model to an independent dataset to evaluate the performance of a set of selected markers. In particular, we first estimate the predicted risk for each multi-locus genotype in the testing dataset. Based on the order of the multi-locus genotypes from the training dataset and newly estimated predicted riks, we can obtain the predictiveness curve, which is not necessary monotonically increasing. A new preidctiveness U can be calculated to evaluate the prediction performance.

For case-control study, we need to first estimate $\hat{P}(g_{i}|D)$
and $\hat{P}(g_{i}|\bar{D})$ from cases and controls seprately. Then,
using $\rho$ estimated from another cohort study, we can obtain $\hat{r}_{i}$,
$\hat{p_{i}}$, and thus the predictiveness U, $\hat{U}=2\sum_{i>j}\hat{p}_{i}\hat{p}_{j}(\hat{r}_{i}-\hat{r}_{j})$.
We can use permutation method to perform hypothesis test and boostrap method to construct confidence interval for predictiveness U. Here, we also
derive the asymptotic variance of predictiveness U. For case-control
data, the predictiveness U can be written as a two sample U statistic (Appendix C),
\[
\hat{U}=2\rho(1-\rho)\frac{1}{n_{D}n_{\bar{D}}}\sum_{s=1}^{n_{D}}\sum_{t=1}^{n_{\bar{D}}}\phi(g_{s},g_{t}),
\]
where $g_{s}$ is the genotype of $s$-th subject in cases, $g_{t}$
is the genotype of $t$-th subject in controls and the kernel $\phi(g_{s},g_{t})$
is defined as, 
\[
\phi(g_{s},g_{t})=\begin{cases}
1, & g_{s}>g_{t}\\
0, & g_{s}=g_{t}\\
-1, & g_{s}<g_{t}
\end{cases}.
\]
The asymptotic variance of predictiveness U can be calculated by (Appendix C),

\begin{align*}
var(\hat{U}) & =4\rho^{2}(1-\rho)^{2}[\frac{1}{n_{D}(n_{D}-1)}\sum_{s=1}^{n_{D}}(\frac{1}{n_{\bar{D}}}\sum_{t=1}^{n_{\bar{D}}}\phi(g_{s},g_{t})-\hat{U})^{2}\\
 & +\frac{1}{n_{\bar{D}}(n_{\bar{D}}-1)}\sum_{t=1}^{n_{\bar{D}}}(\frac{1}{n_{D}}\sum_{s=1}^{n_{D}}\phi(g_{s},g_{t})-\hat{U})^{2}].
\end{align*}

\section*{Result}

We conducted simulation studies and real data analysis to evaluate
the performance of predictiveness U. In particular, we first
compared the performance of standardized global predictiveness U with three other
indexes: Average Entropy ($AE=\int r(x)logr(x)+(1-r(x))log(1-r(x))dx$),
R square ($R=\int(r(x)-\rho)^{2}dx$) and Total Gain ($TG=\int|r(x)-\rho|dx$). To evaluate the performance of
predictiveness U on a small group of individuals carrying rare variants, we compared the
standardized global predictiveness U with the standardized partial predictiveness U.
Finally, we performed a real data application and built a Nicotine Dependence risk
prediction model based on 37 candidate SNPs and 7 environmental risk factors.

\subsection*{Simulation I}

\begin{figure}
\protect\protect\caption{Predictiveness Curves for Simulation I}
\label{Fig2}

\centering \includegraphics[scale=0.7]{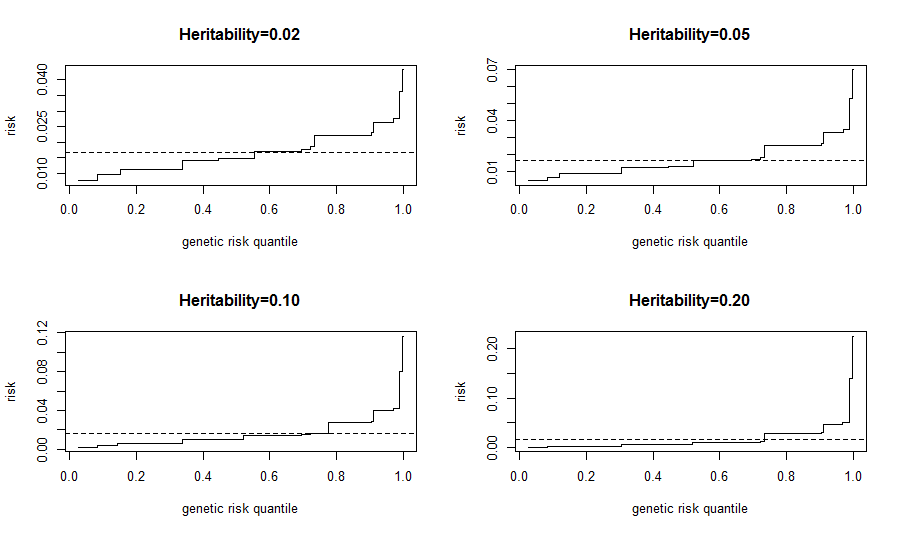}
\end{figure}

We simulated a population with 1 million people with the disease prevalence fixed at 0.016. The 
simulated disease status is influenced by the combining effects of 4 
SNPs (minor allele frequency $> 0.01$), whereby SNPs were assigned with
various marginal effects (additive, dominance or recessive model)
as well as interaction effects among them. We simulated 4 disease
models, by gradually increasing the heritability of the disease (Figure
\ref{Fig2}). We evaluated
the performance of 4 different summary indexes on 1000 case-control data sets sampled from
the simulated population.

The mean and standard deviations for each index were calculated from 1000 data replicates to measure the accuracy and variation of the estimation.
However, since the scale of the indices were different, it was not
convenient to compare their performance based on mean and standard
deviations. Therefore, we calculate \% of Bias and \% of Coverage
to measure the accuracy and variation of the estimator. \% of Bias
is calculated by taking the ratio of the bias and true value. \% of
Coverage is calculated by counting the percentage of the times that
95\% CI covers the true value, whereby the confidence intervals were
calculated by using bootstrap sampling method.

The simulation results are summarized in Table \ref{Tab1}. Overall,
the Bias of predictiveness U is minimal across the 4 summary indices, except that
under high heritability model, TG has comparative bias
as predictiveness U. For \% of coverage,
we found similar trend. As the heritability increase, $TG$, $R$
and $AE$ gain limited improvement, while, predictiveness U consistently
obtained the high coverage rate.

Predictiveness U takes the order of multi-locus genotypes into consideration,
and therefore can handle the non-monotone of the predictiveness
curve when applying model to the testing data. This can partially
explain the advantage of Predictiveness U over the other three indices.
To make the comparison more ``fair'' , we use isotonic
regression to refit the predictiveness curve before summarizing the
curve. By doing so, we force the predictiveness curve to be monotone
increasing in testing data. The new simulation results are summarized in Table \ref{Tab2}.
We found that predictiveness U still has the minimal bias across the
4 indices. In terms of coverage, predictiveness U is comparative with
TG, and much higher than R and AE.

\begin{table}
\protect\protect\caption{Comparison of predictiveness U with three other prediction summary
indices}
\label{Tab1}

\centering %
\begin{tabular}{cccccc}
\hline 
Heritability  &  & U  & R  & TG  & AE\tabularnewline
\hline 
\hline 
0.02  & True value  & 0.199  & 0.00223  & 0.142  & 0.0125\tabularnewline
 & Mean  & 0.194  & 0.00666  & 0.175  & 0.0225\tabularnewline
 & SD  & 0.0173  & 0.00233  & 0.0126  & 0.00308\tabularnewline
 & \% of Bias  & 2.51  & 198  & 22.9  & 79.9\tabularnewline
 & \% of Coverage  & 93.4  & 0  & 0  & 0\tabularnewline
\hline 
0.05  & True value  & 0.312  & 0.00611  & 0.225  & 0.0317\tabularnewline
 & Mean  & 0.305  & 0.0128  & 0.248  & 0.0427\tabularnewline
 & SD  & 0.0178  & 0.00373  & 0.0137  & 0.00483\tabularnewline
 & \% of Bias  & 2.26  & 109  & 10.4  & 34.8\tabularnewline
 & \% of Coverage  & 90.3  & 0  & 8.8  & 0\tabularnewline
\hline 
0.1  & True value  & 0.437  & 0.0145  & 0.328  & 0.0654\tabularnewline
 & Mean  & 0.436  & 0.0244  & 0.346  & 0.0791\tabularnewline
 & SD  & 0.0161  & 0.00553  & 0.0138  & 0.00666\tabularnewline
 & \% of Bias  & 0.14  & 67.8  & 5.31  & 21\tabularnewline
 & \% of Coverage  & 92.6  & 0.2  & 34.5  & 11\tabularnewline
\hline 
0.2  & True value  & 0.599  & 0.041  & 0.475  & 0.138\tabularnewline
 & Mean  & 0.596  & 0.0576  & 0.476  & 0.153\tabularnewline
 & SD  & 0.0138  & 0.0108  & 0.0138  & 0.01\tabularnewline
 & \% of Bias  & 0.464  & 40.5  & 0.173  & 11.2\tabularnewline
 & \% of Coverage  & 92.8  & 0.066  & 86.2  & 11.5\tabularnewline
\hline 
\end{tabular}
\end{table}

\begin{table}
\protect\protect\caption{Comparison of predictiveness U with three other prediction summary
indices under monotonic restriction}
\label{Tab2}

\centering %
\begin{tabular}{cccccc}
\hline 
Heritability  &  & U  & R  & TG  & AE\tabularnewline
\hline 
\hline 
0.02  & True value  & 0.199  & 0.00224  & 0.142  & 0.0125\tabularnewline
 & Mean  & 0.202  & 0.00302  & 0.146  & 0.0141\tabularnewline
 & SD  & 0.0168  & 0.00103  & 0.014  & 0.00237\tabularnewline
 & \% of Bias  & 1.39  & 35  & 2.29  & 12.9\tabularnewline
 & \% of Coverage  & 92.3  & 60.8  & 90.9  & 73.8\tabularnewline
\hline 
0.05  & True value  & 0.312  & 0.00612  & 0.225  & 0.0317\tabularnewline
 & Mean  & 0.312  & 0.00795  & 0.228  & 0.0337\tabularnewline
 & SD  & 0.0175  & 0.00227  & 0.0149  & 0.00415\tabularnewline
 & \% of Bias  & 0.006  & 30  & 1.39  & 6.42\tabularnewline
 & \% of Coverage  & 91.9  & 55.1  & 90.8  & 78.7\tabularnewline
\hline 
0.1  & True value  & 0.437  & 0.0145  & 0.328  & 0.0654\tabularnewline
 & Mean  & 0.442  & 0.0179  & 0.334  & 0.0694\tabularnewline
 & SD  & 0.0159  & 0.00387  & 0.0146  & 0.00597\tabularnewline
 & \% of Bias  & 1.29  & 23.2  & 1.75  & 6.13\tabularnewline
 & \% of Coverage  & 88.1  & 57.6  & 90.1  & 74.2\tabularnewline
\hline 
0.2  & True value  & 0.599  & 0.041  & 0.475  & 0.138\tabularnewline
 & Mean  & 0.601  & 0.0476  & 0.471  & 0.142\tabularnewline
 & SD  & 0.0136  & 0.00864  & 0.0142  & 0.00906\tabularnewline
 & \% of Bias  & 0.399  & 15.9  & 0.759  & 3.25\tabularnewline
 & \% of Coverage  & 91.1  & 67.9  & 93.2  & 81.3\tabularnewline
\hline 
\end{tabular}
\end{table}

\begin{figure}
\protect\protect\caption{Predictiveness Curve for Simulation II}
\label{Fig3}\centering

\includegraphics[scale=0.60]{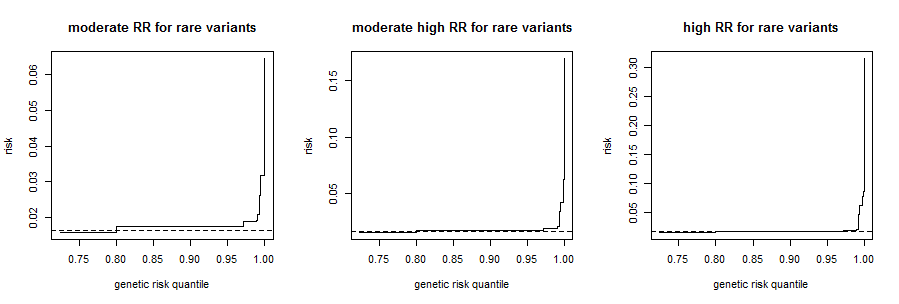} 
\end{figure}

\subsection*{Simulation II}

Clinical interest sometimes focuses on a specific risk quantile
rather than the whole population. In such a case, partial predictiveness U
is more useful than global predictiveness U. Here, we simulated
three disease models, each influenced by 2 common
variants and 2 rare variants. We fixed the effects of the common variants at moderate level while varying the effects of rare variants 
from moderate level to high level (Figure \ref{Fig3}). From Figure \ref{Fig3} we observed that the rare variants
mainly affected the high risk subpopulation.
We use partial predictiveness U to summarize the predictiveness curve
on the highest 10\% risk quantile. Similarly as in
simulation I, we investigated the performance of the two indices, using 1000 sampled data sets from simulated population under each disease model. The simulation results are
summarized in Table \ref{Tab3}. We found that, as the effect size
of rare variants increases, both global predictiveness U and partial
predictiveness U have better prediction. Partial predictiveness
U outperformed global predictiveness U in
terms of both \% of Bias and \% of Coverage. This is probably due to the fact
that partial predictiveness U focuses on the high risk subpopulation, and for the three rare-variants-dominated disease models, the high risk quantile summarized majority of the information, which is in favor of
partial predictiveness U.

\begin{table}
\protect\protect\caption{Comparison of predictiveness U and partial predictiveness U}
\label{Tab3}\centering

\begin{tabular}{ccccccc}
\hline 
 & \multicolumn{2}{c}{moderate $RR_{rare}$} & \multicolumn{2}{c}{moderate high $RR_{rare}$} & \multicolumn{2}{c}{high $RR_{rare}$}\tabularnewline
 & $U$  & $U_{pt}$  & $U$  & $U_{pt}$  & $U$  & $U_{pt}$\tabularnewline
\hline 
\hline 
True value  & 0.0298  & 0.103  & 0.0432  & 0.243  & 0.0527  & 0.312\tabularnewline
Mean  & 0.0201  & 0.13  & 0.0325  & 0.251  & 0.0422  & 0.316\tabularnewline
SD  & 0.0147  & 0.061  & 0.0142  & 0.0563  & 0.0147  & 0.0537\tabularnewline
\%Bias  & 32.6  & 26.4  & 24.7  & 3.3  & 20  & 1.38\tabularnewline
\%Coverage  & 32.3  & 58.8  & 64.8  & 98.4  & 83.3  & 100\tabularnewline
\hline 
\end{tabular}
\end{table}

\subsection*{Application to Nicotine Dependence}

Tobacco-attributable morbidity and mortality represents a remarkable 21st century global
burden of disease. In US population, the lifetime prevalence of nicotine dependence is
about 24\%. Associated with cigarette smoking, it becomes a leading
risk factor for lung cancer. Both gene and environment play an important role
in nicotine dependence. Twin and family studies estimated
that the heritability of nicotine dependence to be around 59\%.\cite{Li2003}
Despite the high heritability, the currently-discovered SNPs has limited value in predicting Nicotine Dependence. In this application, we combined genetic
variants and environmental determinants to build a
prediction model for Nicotine dependence, using a GWAS dataset from the study of Addiction: Genetics and Environment (SAGE).

Two independent samples in SAGE: Family Study of Cocaine dependence
(FSCD) and Collaborative Genetic Study of Nicotine Dependence (COGEND),
were used as training data and validation data respectively. DSM-IV
defined Nicotine Dependence was set as binary outcome for analysis.
In particular, there are 486 cases and 684 controls in FSCD and 696
cases and 708 controls in COGEND. From previous literature, we selected
37 SNPs that were potentially associated with Nicotine dependence.
7 non-genetic risk factors, including sex, age, education, income,
and 3 types of trauma experience were also included in the analysis.

\begin{figure}
\protect\protect\caption{Predictiveness curves as model complexities changes}

\label{Fig4}

\centering \includegraphics{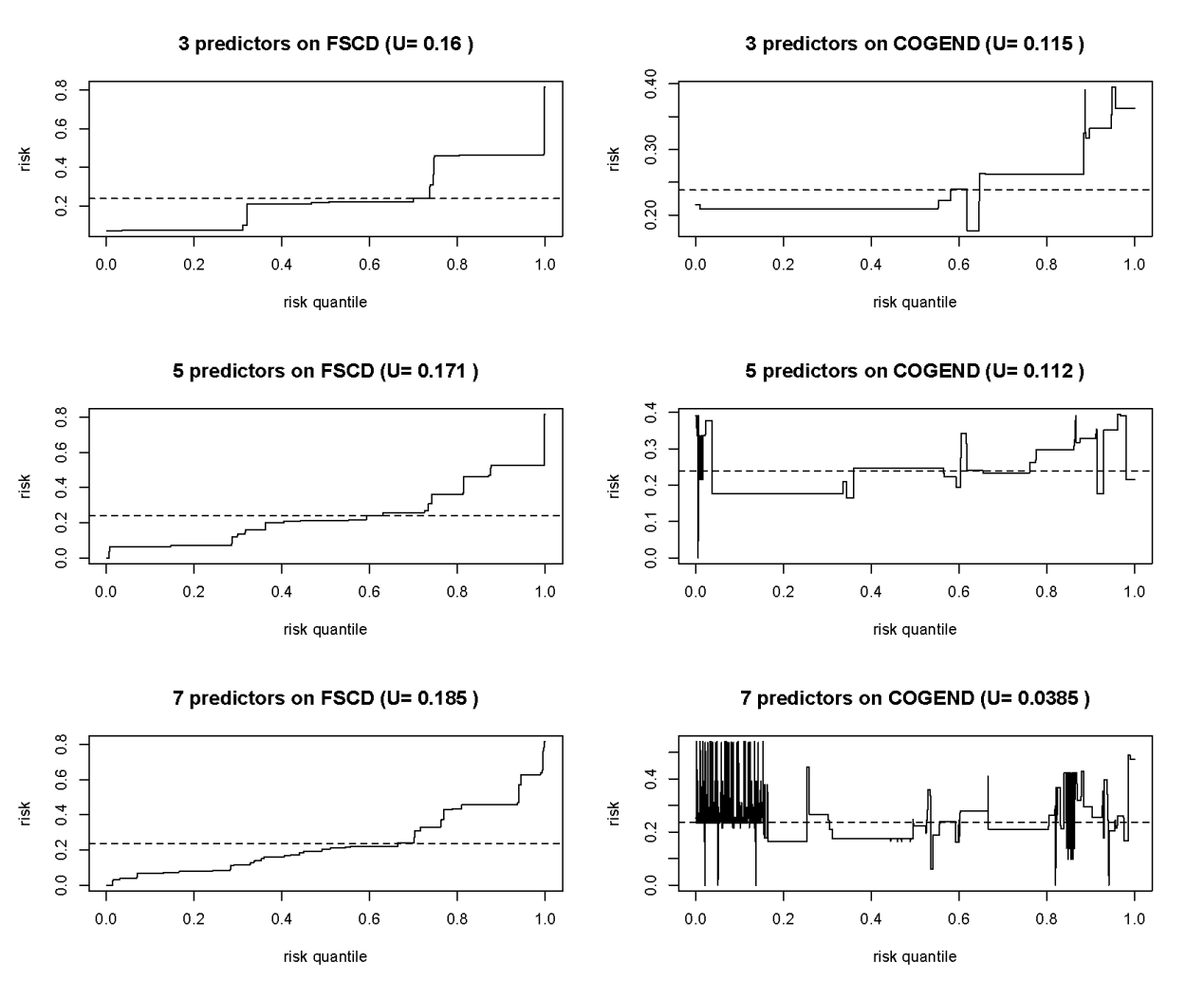}
\end{figure}

To determine a
parsimonious model, we applied a forward selection algorithm\cite{Ye2011} with a build-in cross-validation process to the 44 predictors on the training data. The algorithm selected 3 risk predictors: income, physical trauma and a genetic variant, rs2656073, into the final risk prediction model. We then visualized the prediction
model by a predictiveness curve. The
curve showed moderate prediction accuracy, with a global predictiveness
U of 0.160. We further evaluated the performance of the prediction model using the COGEND dataset. The predictiveness curve on the validation data showed similar patterns as the curve on training data, with a predictiveness U of 0.115 (Figure \ref{Fig4}).

\begin{figure}
\protect\protect\caption{Comparison of Predictiveness U and other summary indices as model
complexiity changes}
\label{Fig5}

\centering \includegraphics[scale=0.75]{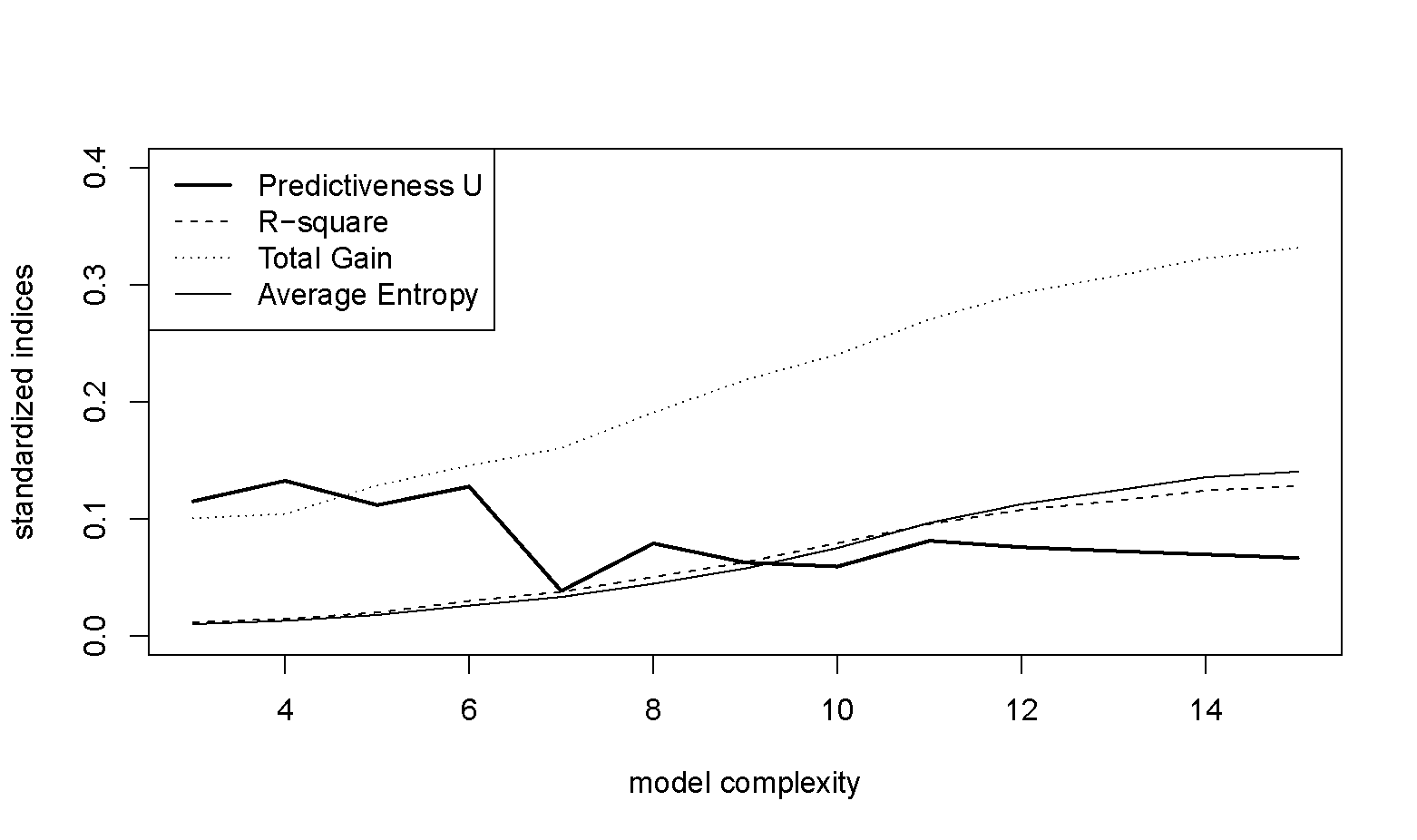}
\end{figure}

We sequentially added 2 predictors into the optimal prediction model
on the forward selection path (Figure \ref{Fig4}). Consistent with
the 10-fold cross validation results, the prediction performance on
COGEND data decreased as number of predictor was larger than 3. For
5-preidctors model, the predictiveness curve on COGEND data still
maintained increasing trend (U=0.112), yet, with a lot of irregularity.
For 7-predictors model, we can hardly observed any increasing trend (U=0.0385).
As the model complexity is above the optimal threshold,
the prediction performance on the testing data decreases due
to over-fitting. We plotted value of the predictiveness U and 3 other summary indices (R, AE, TG) as the number of model complexity
increased (Figure \ref{Fig5}). Predictiveness U showed decreasing
trend with some vibration, which reflected the overfitting of the model. Yet, the other 3 indices R, AE and TG show
consistently increasing trend, contradicting with the reults in Figure \ref{Fig4}.

\section*{Discussion}

The translation of human genome discoveries into health practice is one of the major challenges in
the coming decades. The use of emerging genetic findings from ongoing genome-wide association
studies and next-generation sequencing studies for early disease prediction, prevention and
pharmacogenetics will advance genome medicine and lead to more effective personalized
prevention/treatment strategies\cite{Janssens2011}.
A predicted risk score is helpful for physicians to make personalized
clinical decisions. Meanwhile, to evaluate the population impact of a risk model,
one needs to consider the distribution of the predictors. Yet, ROC
curve, the widely used tool for evaluating prediction model, has recently
been criticized for its limitation to reflect population level information\cite{Cook2007,Pepe2008a,Gail2005}.
Predictiveness curve, on the other hand, can access the usefulness
of a risk prediction model on the population level. In this paper,
we proposed an approach to construct predictiveness curve
for multiple risk predictors. We also provided two
summary statistics for predictiveness curve, a global predictiveness U and a particial
predictiveness U, to assess the predictive value of the model. A global predictiveness
U can be calculated to evaluate the prediction performance across the whole population,
while a partial predictiveness U is proposed for the clinical
interests on a subpopulation (e.g. high risk subpopulation).

In this study, we used the risk difference kernel for predictiveness
U, and compare our approach with 3 other popular summary indices:
R square, Average Entropy and Total Gain. Thorough simulation studies, we showed that R, AE and TG were biased, especially when
the heritability is moderate or low, while Predictiveness U obtained
robust and accurate estimation. This is partly due to the fact that Predictiveness U can capture the monotonicity of the multi-locus predictiveness curve. Predictiveness curve was first introduced to evaluate the predictive value of a single continuous biomarker with implicit underlying order. Therefore, it's convenient to construct a monotone increasing preidctiveness curve. However, for categorical multi-locus genetic marker, the extension is not straightforward. Since the order of the multi-locus markers depends on the underlying
disease model, the predictiveness curve might not monotonically increase in the validation data. Predictiveness U has the capacity to take into account the underlying order of different multilocus genotype and capture the non-monotone information, and thus the advantage over the other summary indices. Although one can use isotonic regression to smooth the
curve and force it to be monotone increasing, this
additional smoothing procedure is subject to information loss. Nevertheless, to make the
comparison more ``fair'', we performed a simulation by adding the smoothing procedure
before calculating summary statistics. Both TG and Predictiveness U had satisfactory
results, while R and AE still obtained biased and unstable estimation.

Predictiveness U is built under the framework of U statistic. U statistic,
first introduced by Hoeffding\cite{Hoeffding1948}, is well suitable for
nonparametric analysis for high dimension data and have been widely
used in genetic association study recently. We here employ this idea
in risk prediction study and proposed a preidctiveness U to measure
the preidctiveness of a set of markers. Predictiveness U is unbiased
estimator of the kernel $\psi(r_{i},r_{j})$. Taking the risk difference
kernel $\psi(r_{i},r_{j})=r_{i}-r_{j}$, predictiveness U can be interpreted
as the expected risk difference across the whole population. If a
set of marker can predict the disease status well, it's reasonable
to see the risk difference between high risk group and low risk group to be large.
Using the standardized form, with scale from 0 to 1, it measures the
proportion of risk difference explained by current model. By taking the underlying order of genetic variants taken into account, it can also serve as
goodness of fit for the population prediction model.

Although predictiveness U aims at evaluating prediction model at population
level, one should always be cautious at how to interpret the result.
A high predictiveness U value obtained in one population does not always mean that the model
will predict as accurately in another population. Disease-associated genetic markers in different population might
be associated with different effect size. Even with the same effect size, the allele or genotype
distribution of markers might also be different, so as to infuence the final results. Therefore,
one should expect to see different predictiveness U in two heterogeneous populations. Furthermore, a low predictiveness
U does not necessarily mean the set of genetic variants is useless.
Sometimes, the clinical interest is on the high risk subpopulation.
For example, some rare variants might not influence the overall risk
in the whole population, but has a large effect on the high risk subpopulation.
A partial predictiveness U can reflect the predictiveness of the set
of markers on specific risk quantile. Through simulation, we showed
that when the rare variants dominated in the disease model, partial
predictiveness U is more suitable than predictiveness U on evaluation
the prediction model.

The primary study design that predictiveness U can be applied is cohort
design, where the population level information is well presented in the
study. With the knowledge of disease prevalence, predictiveness U
can be easily extended to case-control or cross-sectional design.
In our real data application, we analyzed
FSCD and COGEND case-control dataset by using disease prevalence of
24\%. Yet, One
should be cautious to infer population predictiveness from
case-control studies. The study samples might not be representative
for the whole population, due to eligibility criteria and sampling
variability. Furthermore, population substructure and correlation
between subjects can lead to false positive results. Different studies on the same population might end with different prediction model due to heterogeneity in studies (e.g. different genetic array and quality control process).

In the general formula of predictiveness U, we provided a flexible
risk kernel. The choice of the kernel depends on the research
questions to be answered. For different diseases, the cost associated
with the disease risk can vary. One might want to give a heavy weight
on high risk group if it has severe impact. Even for the same disease, the
impact can vary according to the characteristics of the population
(e.g. medical and economic condition). The information can be incorporated
in the risk kernel if needed. In this paper, we attempted to build
a risk prediction model for Nicotine Dependence. The primary interest
here was to find optimal genetic model that can differentiate the
high risk group with low risk group in the population. Therefore, we used risk difference kernel.

We included 37 previously reported SNPs to build the risk prediction
model for Nicotine Dependence. To increase the prediction power, we
also include several environmental risk factors. The optimal model
selected 3 predictors: income, physical trauma and a genetic variant:
rs2656073, which attained a moderate predictiveness U value on the testing data. 
Further investigation of the final model
revealed that the genetic variant, compared with the environmental risk factors, only had a
small effect on the overall prediction. This is reasonable as for psychiatric diseases individual
variants usually have small marginal effects. We can
incorporate an ensemble approach to combine multiple small effect genetic variants for future extension. Due to
the limitations of available information in the dataset, only 7 environmental risk
predictors were included in the analysis. A better predictiveness U model can be
attained by including more comprehensive environmental risk predictors.

\section*{Appendix}

\subsection*{Appendix A}

Let $F(g)$, $F_{D}(g)$ and $F_{\bar{D}}(g)$ denote the c.d.f of
ordered genotype $g$ in the whole population, disease population and non-disease population.
We know for ROC curve, $t_g=1-F_{\bar{D}}(g)$ and $f(t_g)=1-F_{D}(g).$
Let $F_{D}^{\prime}(g)$ and $F_{\bar{D}}^{\prime}(g)$ be the p.d.f
of $g$ in the case and control distribution, we have $f^{\prime}(t)=\frac{df(t)}{dt}=\frac{F_{D}^{\prime}(g)}{F_{\bar{D}}^{\prime}(g)}$.
Then each point $(q,r(q))$ on the predictiveness curve can be written
as,

\begin{align*}
q & =F(g)\\
 & =\rho F_{D}(g)+(1-\rho)F_{\bar{D}}(g)\\
 & =\rho(1-f(t))+(1-\rho)(1-t)
\end{align*}
and 
\begin{align*}
r(q) & =P(D|g)\\
 & =\frac{P(g|D)P(D)}{P(g|D)P(D)+P(g|\bar{D)}P(\bar{D)}}\\
 & =\frac{F_{D}^{\prime}(g)\rho}{F_{D}^{\prime}(g)\rho+F_{\bar{D}}^{\prime}(g)(1-\rho)}\\
 & =\frac{f^{\prime}(t)\rho}{f^{\prime}(t)\rho+(1-\rho)}
\end{align*}

Now we aim to express $U=2\int_{0}^{1}\int_{0}^{y}(r(y)-r(x))dxdy$
in the form of $f(\cdot)$. Let $x=\rho(1-f(t))+(1-\rho)(1-t)$ and
$y=\rho(1-f(s)+(1-\rho)(1-s)$. We can calculate dirivarive of $x$,
\[
\frac{dx}{dt}=-(1-\rho)-\rho f^{\prime}(t)
\]
Since $f(0)=0$ and $f(1)=1$, we know 
\[
x=1\Leftrightarrow t=0,
\]
and

\[
x=0\Leftrightarrow t=1.
\]
Then, we can show 
\begin{align*}
U & =2\int_{0}^{1}\int_{0}^{y}(r(y)-r(x))dxdy\\
 & =2\int_{1}^{0}\int_{1}^{s}(\frac{f^{\prime}(s)\rho}{f^{\prime}(s)\rho+(1-\rho)}-\frac{f^{\prime}(t)\rho}{f^{\prime}(t)\rho+(1-\rho)})[(1-\rho)+\rho f^{\prime}(t)][(1-\rho)+\rho f^{\prime}(s)]dtds\\
 & =2\int_{0}^{1}f^{\prime}(s)\rho\int_{s}^{1}[(1-\rho)+\rho f^{\prime}(t)]dtds-2\int_{0}^{1}[(1-\rho)+\rho f^{\prime}(s)]\int_{s}^{1}f^{\prime}(t)\rho dtds\\
 & =2\int_{0}^{1}f^{\prime}(s)\rho[(1-\rho)(1-s)+\rho(1-f(s))]ds-2\int_{0}^{1}[(1-\rho)+\rho f^{\prime}(s)](\rho-\rho f(s))ds\\
 & =2\rho(1-\rho)\int_{0}^{1}[f^{\prime}(s)-sf^{\prime}(s)-1+f(s)]ds
\end{align*}
Additinally, because $\int_{0}^{1}f^{\prime}(s)ds=1$ and $\int_{0}^{1}sf^{\prime}(s)ds=sf(s)|_{0}^{1}-\int_{0}^{1}f(s)ds$,
the above can be further simplified as,

\[
U=2\rho(1-\rho)[2\int_{0}^{1}f(s)ds-1]=2\rho(1-\rho)(2AUC_{R}-1).
\]

\subsection*{Appendix B}

We will show the relationship between predictiveness U and $AUC_{L}$
by first showing the relationship between $AUC_{R}$ and $AUC_{L}$.
Using notation in Appendix A and definition of Lorenze curve, 
\begin{align*}
AUC_{L} & =\frac{1}{\rho}\int_{0}^{1}\int_{0}^{y}r(x)dxdy\\
 & =\int_{0}^{1}\int_{s}^{1}f^{\prime}(t)dt[1-\rho+\rho f^{\prime}(s)]ds\\
 & =\int_{0}^{1}[1-f(s)](1-\rho+\rho f^{\prime}(s))ds\\
 & =(1-\rho)(1-\int_{0}^{1}f(s)ds)+\rho\int_{0}^{1}f^{\prime}(s)ds-\rho\int_{0}^{1}f(s)f^{\prime}(s)ds\\
 & =(1-\rho)(1-AUC_{R})+\rho-\rho\int_{0}^{1}f(s)f^{\prime}(s)ds
\end{align*}
Since $\int_{0}^{1}f(s)f^{\prime}(s)ds=f(s)f(s)|_{0}^{1}-\int_{0}^{1}f^{\prime}(s)f(s)ds$,
we know $\int_{0}^{1}f(s)f^{\prime}(s)ds=\frac{1}{2}$. Thus, the
above equation can be simplified, 
\[
AUC_{L}=(1-\rho)(1-AUC_{R})+\frac{1}{2}\rho.
\]

Since $U=2\rho(1-\rho)(2AUC_{R}-1)$, we have 
\[
U=2\rho(0.5-AUC_{L}).
\]

\subsection*{Appendix C}

The estimator of predictiveness U can be written as $\hat{U}=2\sum_{i>j}\hat{p}_{i}\hat{p}_{j}(\hat{r}_{i}-\hat{r}_{j})$,
where, we can calculate $\hat{p}_{i}$ and $\hat{r}_{i}$ from $\hat{P}(g_{i}|D$)
and $\hat{P}(g_{i}|\bar{D})$. Thus we can also write predictiveness
U as function of $\hat{P}(g_{i}|D$) and $\hat{P}(g_{i}|\bar{D})$.
\begin{align*}
\hat{U} & =2\sum_{i=1}^{G_{m}}\sum_{j=1}^{i}\hat{p}_{i}\hat{p}_{j}(\hat{r}_{i}-\hat{r}_{j})\\
 & =2\sum_{i=1}^{G_{m}}\hat{P}(g_{i},D)\sum_{j=1}^{i}\hat{P}(g_{i})-2\sum_{i=1}^{G_{m}}\hat{P}(g_{i})\sum_{j=1}^{i}\hat{P}(g_{j},D)\\
 & =2\sum_{i=1}^{G_{m}}\rho\hat{P}(g_{i}|D)\sum_{j=1}^{i}[\rho\hat{P}(g_{j}|D)+(1-\rho)\hat{P}(g_{j}|\bar{D})]\\
 & -2\sum_{i=1}^{G_{m}}[\rho\hat{P}(g_{i}|D)+(1-\rho)\hat{P}(g_{i}|\bar{D})]\sum_{j=1}^{i}\rho\hat{P}(g_{j}|D)\\
 & =2\rho(1-\rho)[\sum_{i=1}^{G_{m}}\hat{P}(g_{i}|D)\sum_{j=1}^{i}\hat{P}(g_{j}|\bar{D})-\sum_{i=1}^{G_{m}}\hat{P}(g_{i}|\bar{D})\sum_{j=1}^{i}\hat{P}(g_{j}|D)].
\end{align*}
Observing the similarities of the two summations, we can write the
above equation as 
\[
\hat{U}=2\rho(1-\rho)[\sum_{i=1}^{G_{m}}\sum_{j=1}^{G_{m}}\hat{P}(g_{i}|D)\hat{P}(g_{j}|\bar{D})(\mathbf{1}_{\{i>j\}}-\mathbf{1}_{\{i<j\}})],
\]
where, $\mathbf{1}_{\{\cdot\}}$ is indicator function. Pluging in
the estimator $\hat{P}(g_{i}|D)=\frac{n_{g_{i},D}}{n_{D}}$ and $\hat{P}(g_{j}|\bar{D})=\frac{n_{g_{i},\bar{D}}}{n_{\bar{D}}}$,
we can show that the predictiveness U is a two sample U statistics,
\begin{align*}
\hat{U} & =2\rho(1-\rho)\frac{1}{n_{D}n_{\bar{D}}}\sum_{i=1}^{G_{m}}\sum_{j=1}^{G_{m}}n_{g_{i},D}n_{g_{j},\bar{D}}(\mathbf{1}_{\{i>j\}}-\mathbf{1}_{\{i<j\}})\\
 & =2\rho(1-\rho)\frac{1}{n_{D}n_{\bar{D}}}\sum_{s=1}^{n_{D}}\sum_{t=1}^{n_{\bar{D}}}\phi(g_{s},g_{t}),
\end{align*}
where $g_{s}$ is the genotype of $s$-th subject in cases, $g_{t}$
is the genotype of $t$-th subject in controls and the kernel $\phi(g_{s},g_{t})$
is defined as, 
\[
\phi(g_{s},g_{t})=\begin{cases}
1, & g_{s}>g_{t}\\
0, & g_{s}=g_{t}\\
-1, & g_{s}<g_{t}
\end{cases}.
\]
Let $\theta=E(\phi(g_{s},g_{t}))$ and $\theta_{U}=E(\hat{U})=2\rho(1-\rho)\theta$.
We can calculate variance of $\hat{U}-\theta_{U}$, 
\begin{align*}
var(\hat{U}-\theta_{U}) & =\frac{4\rho^{2}(1-\rho)^{2}}{n_{D}^{2}n_{\bar{D}}^{2}}var[\sum_{s=1}^{n_{D}}\sum_{t=1}^{n_{\bar{D}}}(\phi(g_{s},g_{t})-\theta)]\\
 & =\frac{4\rho^{2}(1-\rho)^{2}}{n_{D}^{2}n_{\bar{D}}^{2}}[n_{D}n_{\bar{D}}\tau_{1,1}+n_{D}n_{\bar{D}}(n_{\bar{D}}-1)\tau_{1,0}+n_{D}n_{\bar{D}}(n_{D}-1)\tau_{0,1}],
\end{align*}
where, $\tau_{1,1}=var(\phi(g_{s},g_{t}))$, $\tau_{1,0}=cov(\phi(g_{s},g_{t}),\phi(g_{s},g_{t^{\prime}})$)
and $\tau_{0,1}=cov(\phi(g_{s},g_{t}),\phi(g_{s^{\prime}},g_{t}))$.

To obtain the asymtotic distribution of $\hat{U}$, we can use Hajek
projection to project $\hat{U}-\theta_{U}$ onto the space of the
summation forms $\sum_{k=1}^{n_{D}+n_{\bar{D}}}h(g_{k})$, where the
CLT can be easily applied. The Hajek projection $\tilde{U}$ of $\hat{U}-\theta_{U}$
is, 
\begin{align*}
\tilde{U} & =\sum_{s=1}^{n_{D}}E(\hat{U}-\theta_{U}|g_{s})+\sum_{t=1}^{n_{\bar{D}}}E(\hat{U}-\theta_{U}|g_{t})\\
 & =\frac{2\rho(1-\rho)}{n_{D}}\sum_{s=1}^{n_{D}}h_{1,0}(g_{s})+\frac{2\rho(1-\rho)}{n_{\bar{D}}}\sum_{s=1}^{n_{\bar{D}}}h_{0,1}(g_{t}),
\end{align*}
where $h_{1,0}(g_{s})=E(\phi(g_{s},g_{t})-\theta|g_{s})$ and $h_{0,1}(g_{t})=E(\phi(g_{s},g_{t})-\theta|g_{t})$.
We can then calculate the variance of $\tilde{U}$, 
\begin{align*}
var(\tilde{U}) & =\frac{4\rho^{2}(1-\rho)^{2}}{n_{D}}var(h_{1,0}(g_{s}))+\frac{4\rho^{2}(1-\rho)^{2}}{n_{\bar{D}}}var(h_{0,1}(g_{t}))\\
 & =4\rho^{2}(1-\rho)^{2}[\frac{\tau_{1,0}}{n_{D}}+\frac{\tau_{0,1}}{n_{\bar{D}}}].
\end{align*}

We can write $\hat{U}-\theta_{U}$ as a summation of the projection
term $\tilde{U}$ and the remainding term $\tilde{R}$, i.e., $\hat{U}-\theta_{U}=\tilde{U}+\tilde{R}$.
The asymptotic normaility of $\hat{U}-\theta_{U}$ is then established
by showing $\tilde{U}$ is asymtotically normal and $\tilde{R}$ is
asymtotically negligiable. Assuming $n=n_{D}+n_{\bar{D}}$, $\frac{n_{D}}{n}\rightarrow\lambda$,
we can apply CLT to $\tilde{U}$ and show that, 
\[
\sqrt{n}\tilde{U}\rightsquigarrow N(0,4\rho^{2}(1-\rho)^{2}[\frac{\tau_{1,0}}{\lambda}+\frac{\tau_{0,1}}{1-\lambda}]).
\]
With the fact that $E(\tilde{U})=0$, $E(\tilde{R})=0$ and $E(\tilde{U}\tilde{R})=0$,
we know $E(n\tilde{R}^{2})=nvar(\hat{U}-\theta)-nvar(\tilde{U})\rightarrow0$.
Thus, $\sqrt{n}\tilde{R}\overset{p}{\rightarrow}0$. With slusky theorem,
we know, 
\[
\sqrt{n}(\hat{U}-\theta)\rightsquigarrow N(0,4\rho^{2}(1-\rho)^{2}[\frac{\tau_{1,0}}{\lambda}+\frac{\tau_{0,1}}{1-\lambda}]).
\]

\bibliography{pdu}

\begin{thebibliography}{10}
\providecommand{\url}[1]{\texttt{#1}}
\providecommand{\urlprefix}{URL }
\expandafter\ifx\csname urlstyle\endcsname\relax
  \providecommand{\doi}[1]{doi:\discretionary{}{}{}#1}\else
  \providecommand{\doi}{doi:\discretionary{}{}{}\begingroup
  \urlstyle{rm}\Url}\fi

\bibitem{Pepe2008}
Pepe MS, Feng Z, Huang Y, Longton G, Prentice R, Thompson IM, Zheng Y.
  Integrating the predictiveness of a marker with its performance as a
  classifier. \emph{American Journal of Epidemiology}  2008;
  \textbf{167}(3):362--368, \doi{10.1093/aje/kwm305}.
  \urlprefix\url{http://aje.oxfordjournals.org/content/167/3/362.abstract}.

\bibitem{Huang2007}
Huang Y, Pepe MS, Feng Z. Evaluating the predictiveness of a continuous marker.
  \emph{Biometrics}  2007; \textbf{63}(4):1181--1188. \urlprefix\url{<Go to
  ISI>://WOS:000251508300023}.

\bibitem{Huang2010}
Huang Y, Pepe MS. Assessing risk prediction models in case-control studies
  using semiparametric and nonparametric methods. \emph{Statistics in Medicine}
   2010; \textbf{29}(13):1391--1410. \urlprefix\url{<Go to
  ISI>://WOS:000279060000006}.

\bibitem{Mittlbock1996}
Mittlbock M, Schemper M. Explained variation for logistic regression.
  \emph{Statistics in Medicine}  1996; \textbf{15}(19):1987--1997.
  \urlprefix\url{<Go to ISI>://WOS:A1996VK32500001}.

\bibitem{Cox1992}
Cox DR, Wermuth N. A comment on the coefficient of determination for binary
  responses. \emph{American Statistician}  1992; \textbf{46}(1):1--4.
  \urlprefix\url{<Go to ISI>://WOS:A1992HB06200001}.

\bibitem{Shapiro1977}
Shapiro AR. Evaluation of clinical predictions - method and initial
  application. \emph{New England Journal of Medicine}  1977;
  \textbf{296}(26):1509--1514. \urlprefix\url{<Go to
  ISI>://WOS:A1977DL21900007}.

\bibitem{Bura2001}
Bura E, Gastwirth JL. The binary regression quantile plot: Assessing the
  importance of predictors in binary regression visually. \emph{Biometrical
  Journal}  2001; \textbf{43}(1):5--21. \urlprefix\url{<Go to
  ISI>://WOS:000167263200002}.

\bibitem{Hoeffding1948}
Hoeffding W. A class of statistics with asymptotically normal distribution.
  \emph{Annals of Mathematical Statistics}  1948; \textbf{19}(3):293--325.
  \urlprefix\url{<Go to ISI>://WOS:A1948UM00800001}.

\bibitem{Wei2008}
Wei Z, Li MY, Rebbeck T, Li HZ. U-statistics-based tests for multiple genes in
  genetic association studies. \emph{Annals of Human Genetics}  2008;
  \textbf{72}:821--833. \urlprefix\url{<Go to ISI>://WOS:000259936700012}.

\bibitem{Schaid2005}
Schaid DJ, McDonnell SK, Hebbring SJ, Cunningham JM, Thibodeau SN.
  Nonparametric tests of association of multiple genes with human disease.
  \emph{American Journal of Human Genetics}  2005; \textbf{76}(5):780--793.
  \urlprefix\url{<Go to ISI>://WOS:000228198300006}.

\bibitem{Lu2008}
Lu Q, Elston RC. Using the optimal receiver operating characteristic curve to
  design a predictive genetic test, exemplified with type 2 diabetes. \emph{The
  American Journal of Human Genetics}  2008; \textbf{82}(3):641--651.
  \urlprefix\url{http://www.sciencedirect.com/science/article/pii/S0002929708001547}.

\bibitem{Li2003}
Li MD, Cheng R, Ma JZ, Swan GE. A meta-analysis of estimated genetic and
  environmental effects on smoking behavior in male and female adult twins.
  \emph{Addiction}  2003; \textbf{98}(1):23--31.
  \urlprefix\url{http://dx.doi.org/10.1046/j.1360-0443.2003.00295.x}.

\bibitem{Ye2011}
Ye CY, Cui YH, Wei CS, Elston RC, Zhu J, Lu Q. A non-parametric method for
  building predictive genetic tests on high-dimensional data. \emph{Human
  Heredity}  2011; \textbf{71}(3):161--170. \urlprefix\url{<Go to
  ISI>://WOS:000293078900003}.

\bibitem{Janssens2011}
Janssens ACJW, Ioannidis JPA, van Duijn CM, Little J, Khoury MJ. Strengthening
  the reporting of genetic risk prediction studies: the grips statement.
  \emph{Eur J Hum Genet}  2011; \textbf{19}(8):833--836.
  \urlprefix\url{http://dx.doi.org/10.1038/ejhg.2011.25
  http://www.nature.com/ejhg/journal/v19/n8/suppinfo/ejhg201125s1.html}.

\bibitem{Cook2007}
Cook NR. Use and misuse of the receiver operating characteristic curve in risk
  prediction. \emph{Circulation}  2007; \textbf{115}(7):928--935,
  \doi{10.1161/circulationaha.106.672402}.
  \urlprefix\url{http://circ.ahajournals.org/content/115/7/928.abstract}.

\bibitem{Pepe2008a}
Pepe MS, Feng Z, Gu JW. Comments on 'evaluating the added predictive ability of
  a new marker: From area under the roc curve to reclassification and beyond'
  by m. j. pencina et al., statistics in medicine. \emph{Statistics in
  Medicine}  2008; \textbf{27}(2):173--181. \urlprefix\url{<Go to
  ISI>://WOS:000253098900002}.

\bibitem{Gail2005}
Gail MH, Pfeiffer RM. On criteria for evaluating models of absolute risk.
  \emph{Biostatistics}  2005; \textbf{6}(2):227--239,
  \doi{10.1093/biostatistics/kxi005}.
  \urlprefix\url{http://biostatistics.oxfordjournals.org/content/6/2/227.abstract}.

\end{thebibliography}
\bibliographystyle{wileyj} 

\end{document}